\documentclass{article}
\usepackage[T1]{fontenc}
\usepackage[utf8]{inputenc}
\usepackage{amsmath,amssymb}
\usepackage[english]{babel}
\usepackage{url}
\usepackage{hyperref}
\usepackage{xcolor}
\begin{document}
\begin{center}
\large{Functional variant of Polynomial Analogue\\ of Gandy's Fixed Point Theorem}
\end{center}
\begin{center}
Andrey Nechesov\\
Sobolev Insitute of Mathematics, Novosibirsk\\
nechesoff@gmail.com
\end{center}
\textbf{Abstact} In this work, a functional variant of the polynomial analogue of the classical Gandy's fixed point theorem is obtained. Sufficient conditions have been found to ensure that the complexity of the recursive function does not go beyond the polynomial.\\

\textbf{Introduction} In 2021, we proved a polynomial analogue of the classical Gandy's fixed point theorem \cite{b_gandy}. This became an important impetus for the construction of p-complete programming languages. And such a language was first built by us in 2022. The main result of that work was: a solution of the problem $P=L$. \cite{b_gn} Next are the followers of the works on building a new high-level language and the idea of building a general programming methodology. But there was one gap in our research: classes of recursive functions whose complexity was polynomial were not described.\\

In this work we found sufficient conditions for such functions. In many ways, the main ideas of this work are similar to the ideas that we used in the proof of the polynomial analogue of Gandy's fixed point theorem. But there are also striking differences. Functions, as such, differ quite strongly from predicates precisely in the multitude of their values. If a predicate is either true or false, then a function can generally take on a variety of values.\\

Moreover, even if there are not many values, but there is recursion and simple multiplication, then powers and factorials may arise during the calculations, which, of course, can violate the polynomial computational complexity of this function. Therefore, finding these restrictions on recursive functions that would be soft enough for the class of functions to be large, and at the same time tough enough not to go beyond polynomiality, has been a problem for us for the last 3 years, after the proof of the polynomial analogue Gandy's fixed point theorem in the case of predicate extensions.\\

In this work, we will inductively expand the function itself, and not the truth set of the predicate, which fundamentally distinguishes these two approaches.\\

Let <HW($\mathfrak{M}$),$\sigma$> -- p-computable hereditary finite list superstructure from [1]. \\
Let $f_1,\dots,f_n$ -- p-computable functions and all $f_i$ in the signature $\sigma$.
Then we can define a new finite or countable generative family of the terms $T_i$ of the signature $\sigma$ for each $f_i$:
\begin{equation}
T_i = \{ t_j(x_1,\dots,x_{k_j})\ |\ j\in N\},\ i\in[1,\dots,n]
\end{equation}
by default we assume that any $f_j$ can be included in any term $t_k$ of any family $T_i$.\\

Let also for each family of terms $T_i$ is defined p-computable function $\gamma_i$:\\
\begin{equation}
\gamma_i:HW(M)\to T_i\cup\{false\}
\end{equation}

By default, we assume that the false element $false\in M$. And in the future, if the function $f(w)$ ($f\in\sigma$) takes the value $false$, then we will denote this as $f(w)\uparrow$, and denote $f(w)\downarrow$ otherwise. The domain of definition $dom(f)$ of a function $f$ will be considered to be all elements $w$ on which $f(w)\downarrow$.\\

Define the extention $f^{(n+1)}_i$ for each $f^{(n)}_i$ (where $f^{(0)}_i=f_i$) and 
\begin{equation}
f^{(n+1)}_i|_{dom(f^{(n)}_i)}=f^{(n)}_i|_{dom(f^{(n)}_i)}
\end{equation}
using a p-computable function $\gamma_i:HW(M)\to T_i\cup\{false\}$ such that on (n+1)-iteration we have:
\begin{equation} \label{eq:fiteration}
f^{(n+1)}_i(w)= \left\{
     \begin{array}{lr}
     f_i(w) \text{, if } f_i(w)\downarrow, \text{ else }\\
     \gamma_i(w)^{\overline{x}}_{\overline{w}} \text{, if } \gamma_i(w)\neq false \text{ and } |\overline{x}| = |\overline{w}|\\               
     false \text{, otherwise }
     \end{array}
     \right.
\end{equation}
and then we are expanding the function $f_i$ to $f^{(n+1)}_i$ and go to the next iteration.\\

Define, $f\subseteq g\Leftrightarrow dom(f)\subseteq dom(g) \text{ and } f|_{dom(f)}=g|_{dom(f)}$.\\ 

Using families of the terms $T_i$, p-computable functions $\gamma_i$ and equality \ref{eq:fiteration} we can define the operator:
\begin{equation}
\Gamma^{<HW(\mathfrak{M}),\sigma>}_{f_1,\dots,f_n}: \{g |\ g:HW(M)\to HW(M)\}^{n} \to \{g | \ g:HW(M)\to HW(M)\}^{n}
\end{equation} 
satisfying the following rules
\begin{equation}
\Gamma^{<HW(\mathfrak{M}),\sigma>}_{f_1,\dots,f_n}(g^{(k)}_1,\dots,g^{(k)}_n) = (g^{(k+1)}_1,\dots,g^{(k+1)}_n)
\end{equation} 
where $g_i^{(k)}\subseteq g_i^{(k+1)}$.\\

The operator $\Gamma^{<HW(\mathfrak{M}),\sigma>}_{f_1,\dots,f_n}$ is monotonic and has the property of a fixed point $(g^{*}_1,\dots,g^{*}_n)$; moreover, the fixed point is reached in $\omega$ steps:

\begin{equation}
\Gamma^{<HW(\mathfrak{M}),\sigma>}_{f_1,\dots,f_n}(f^{*}_1,\dots,f^{*}_n) = (f^{*}_1,\dots,f^{*}_n)
\end{equation}
where $g_i^*=g_i^{(\omega)}$, $i\in[1,\dots,n]$\\

Define a set of free variables of a term  or formula as $V(t(\overline{x}))$ and $V(\varphi(\overline{x}))$ respectively, where $\overline{x}=(x_1,\dots,x_n)$ for some $n\in N$. Denote $V_x(\varphi(\overline{x},\overline{y}))$ and $V_y(\varphi(\overline{x},\overline{y}))$ the set of variable $\{x_i\}$ such that $x_i\in V(\overline{x})$ and $\{y_i\}$ such that $y_i\in V(\overline{y})$. Also denote fpor $t^{\overline{y}}(\overline{x},\overline{y})$ is term where we replaced all occurences $f_j(x_i)$ on $y_i$ in the term $t(\overline{x},\overline{y})$.\\ 

As in the case of the polynomial analogue of Gandy's fixed point theorem, we define the concept of a GNF-system as a tuple:
\begin{equation}\nonumber
 GNF=(\Sigma,\sigma,\Omega, L, HW(M),\Omega,F,T,G)
 \end{equation}

We will say that a GNF-system is a p-computable if the model $HW(\mathfrak{M})$ is a p-computable, all $\gamma_i\in G$ is a p-computable and if the following conditions are met:
\begin{itemize}
	\item if $f_j$ is included in some $t_k$ then only of the form $f_j(x_i)$ for any $x_i$
	\item $f_j(x_i)$ and $f_k(x_i)$ do not belong to any term simultaneosly for any $x_i$
     \item for any $t_j^{\overline{y}}(\overline{x},\overline{y})$ it's not true that there is such $i\in N$ that $x_i\in V(t_j^{\overline{y}}(\overline{x},\overline{y}))$ and $y_i\in V(t_j^{\overline{y}}(\overline{x},\overline{y}))$ 
	\item for any $T_k$ there exists C and p such that for any $t_j\in T_k$ computational complexity $t(t_j^{\overline{y}}(\overline{x},\overline{y}))$ do not exceed $C\cdot (|\overline{x}|+|\overline{y}|)^p$
	\item if $V_y(t_j^{\overline{y}}(\overline{x},\overline{y}))\neq\varnothing$ then the value of the term $|t_j^{\overline{y}}(\overline{w},\overline{l})|$ do not exceed $|\overline{w}|+|\overline{l}|$  for any $t_j\in T_k$, for any evaluates $w_i,l_j\in HW(M)$ of the tuples $\overline{x}$ and $\overline{y}$ respectively.     
\end{itemize}

$\textbf{Theorem (Functional variant of Polynomial Analogue}\\
 \textbf{ of Gandy's Fixed Point Theorem)}:$ Let p-computable GNF-system with p-computable initial functions $f_1,\dots,f_n$ be given, then the smallest fixed point $\Gamma(\overline{f})$ of the operator $\Gamma^{<HW(\mathfrak{M}),\sigma>}_{f_1,\dots,f_n}$ is p-computable.\\

\textbf{Proof:} By induction on the rank $r(w)$ of the element $w$ we prove two statements simultaneously. First,
\begin{equation}
|f_i(w)| \leq C\cdot|w|^p
\end{equation}

and second, that the computational complexity of the algorithm does not exceed the following:
\begin{equation}
t(f_i(w)) \leq 36\cdot C^{p+1}\cdot(r(w)+1)\cdot |w|^{p^2+1}
\end{equation}

Base of induction: $n=0$. Then for some $a\in M$ we have $|f_i(a)|\leq C\cdot|a|^p$ and $t(f_i(a))\leq C\cdot|a|^p$ because initial function $f_i$ is a polynomial computable.\\
Step of induction: Let the statement be true for everyone $k<n$ then show that for $k=n$:

If initial polynomial function $f(w)\downarrow$ then obviosly.\\
Else if $\gamma(w) = false$ then obviously.\\
Else if $\gamma(w) \neq false$ then consider two cases for $I=\{i|y_i\in t^{\overline{y}}_j(\overline{x},\overline{y})\}$:\\
1) $I=\varnothing$ then by conditions for p-computable GNF-system:
\begin{equation}
|f_i(w)| \leq |[\gamma(w)](\overline{w})| = |t_j(w_1,\dots,w_n)|\leq C\cdot|w|^p
\end{equation}
and it is true by definition of p-computable GNF-system.\\

For complexion we have next:
\begin{equation}\nonumber
t(f_i(w))\leq r_{in} + r_{\gamma_i} + t([\gamma_i(w)](\overline{w})) + r_1 = t([\gamma_i(w)^{\overline{y}}](\overline{w}_x,\overline{v}_y)) + r_1 + r_{in} + r_{\gamma_i}\leq 
\end{equation}
\begin{equation}\nonumber
\leq C\cdot(|\overline{w}_x|+|\overline{v}_y|)^p + r_1 +r_{in} + r_{\gamma_i}\leq
\end{equation}
\begin{equation}
\leq C\cdot|\overline{w}|^p + r_1 + r_{in} + r_{\gamma_i} \leq 36\cdot C^{p+1}\cdot(r(w)+1)\cdot |w|^{p^2+1}
\end{equation}
because $\overline{v}$ is empty and where
\begin{itemize}
     \item $r_{in}$ the complexity of checking the value of the initial function $f(w)$, complexity does not exceed $C\cdot|w|^p$
     \item $r_{\gamma_i}$ the complexity of the function $\gamma_i(w)$ for construction the term $t_j(x_1,\dots,x_n)$, complexity does not exceed $C\cdot|w|^p$
     \item $r_1$ complexity of the replacement function which replace $f_j(w_i)$ on $v_i$.
\end{itemize}

2) if $I\neq\varnothing$ then:
\begin{equation}
 |f_i(w)|=|[\gamma_i(w)](\overline{w})| = |[\gamma_i(w)]^{\overline{y}}(\overline{w}_x,\overline{v}_y)|\leq |\overline{w}_x| + |\overline{v}_y| \leq C\cdot|\overline{w}|^p
 \end{equation}
where $v_i = f_{j_i}(w_i)$ for $v_i\in\overline{v}_y$ and $|v_i|\leq C\cdot |w_i|^p$ by induction.

For complexion we have next:

\begin{equation}\nonumber
t(f_i(w))\leq r_{\gamma_i} + t([\gamma_i(w)](\overline{w})) + r_1 \leq t([\gamma_i(w)^{\overline{y}}](\overline{w}_x,\overline{v}_y)) + r_1 + r_2 + r_{\gamma_i} + \sum_{i\in I} t(f(w_i))\leq
\end{equation}
\begin{equation}\nonumber
\leq C\cdot (C\cdot|w|^p)^p +r_1+r_2+ r_{\gamma_i}+\sum_{i=1}^n 36\cdot C^{p+1}\cdot(r(w)-1+1)\cdot |w_i|^{p^2+1}\leq 
\end{equation}
\begin{equation}
\leq 36\cdot C^{p+1}\cdot(r(w)+1)\cdot |w|^{p^2+1}
\end{equation}

\end{document}